\documentclass[12pt]{article}

\usepackage{amssymb,amsmath}
\usepackage{times}
\begin{document}
\setcounter{page}{1}

\newpage
\setcounter{figure}{0}
\setcounter{equation}{0}
\setcounter{footnote}{0}
\setcounter{table}{0}
\setcounter{section}{0}



\centerline{\Large \bf {$p$-Adic  and Adelic Superanalysis} }

\bigskip

\bigskip




\centerline{\large {{Branko
Dragovich}$^{a}$\footnote{dragovich@phy.bg.ac.yu} and {Andrei
Khrennikov}$^{b}$}}
\bigskip

$^a${Institute of Physics, P.O.Box 57, 11001 Belgrade, Serbia and
Montenegro}

\bigskip

 $^b${International Center for Mathematical Modeling,
V\"axj\"o University}

\centerline{V\"axj\"o, Sweden}


\begin{abstract}
After a brief review of $p$-adic numbers, adeles and their
functions, we consider real, $p$-adic and adelic superalgebras,
superspaces and superanalyses. A concrete illustration is given by
means of the Grassmann algebra generated by two anticommuting
elements.
\end{abstract}


\section{Introduction}

Since Einstein's theory of (special and general) relativity, a
symmetry principle has been often taken as a guidance to formulate
a new and more profound physical theory. In the process of gradual
discoveries  more general and fundamental symmetries, a symmetry
called supersymmetry (SUSY) was invented some more than three
decades ago \cite{wess}.

It is well known that SUSY relates basic properties between bosons
and fermions. It plays very important role in construction of new
fundamental models of elementary particle physics beyond the
Standard Model. There are many supersymmetric field theory and
supergravity models. SUSY improves situation with problem of
ultraviolet  divergences, and is important for point particles as
well as for extended objects (strings, branes). In particular,
SUSY plays a significant role in construction of String/M-theory,
which is currently the best candidate for unification of all
interactions and elementary constituents  of matter.

Besides enormous success of SUSY, to  our opinion it should be
extended by the following adelic symmetry principle: a fundamental
physical theory (like String/M-theory) has to be invariant under
some interchange of real and $p$-adic number fields. For the first
time, a similar principle was given by Volovich \cite{volovich}.
There are already some good illustrative examples  of adelic
symmetry in adelic quantum mechanics \cite{dragovich1},
\cite{djordjevic} and adelic string product formulas
\cite{freund2}. To extend SUSY by adelic symmetry it is natural
first to find $p$-adic analogs of standard SUSY (over real
numbers) and then to unify results in the adelic form, which takes
real and all $p$-adic supersymmetries simultaneously and on equal
footing.

In addition to SUSY a strong motivation to consider $p$-adic and
adelic superanalysis  comes also from the quest to formulate
$p$-adic and adelic superstring theory. A notion of $p$-adic
string and hypothesis on the existence of non-archimedean geometry
at the Planck scale were introduced by Volovich \cite{volovich1}
and have been investigated by many researchers (reviews of an
early period are in \cite{vladimirov1} and \cite{brekke}). Very
successful $p$-adic analogs of the Veneziano and Virasoro-Shapiro
amplitudes were proposed in \cite{freund1} as the corresponding
Gel'fand-Graev \cite{gelfand} beta functions. Using this approach,
Freund and Witten obtained \cite{freund2} an attractive adelic
formula $
 A_\infty (a,b) \prod_p A_p (a,b) =1 ,
$ which states that the product of the crossing symmetric
Veneziano (or Virasoro-Shapiro) amplitude and its all $p$-adic
counterparts equals unity (or a definite constant). This gives
possibility to consider an ordinary four-point function, which is
rather complicate (a special function), as an infinite product of
its inverse $p$-adic analogs, which are elementary functions. The
ordinary crossing symmetric Veneziano amplitude can be defined by
a few equivalent ways and its integral form is
\begin{equation}
  A_\infty(a,b)        =  \int_{{\mathbb R}} \vert x \vert_\infty^{a-1}
  \, \vert 1-x\vert_\infty^{b-1} \, dx ,
     \label{1.1}
\end{equation}
where it is taken $\hbar=1,\ T=1/\pi$, and $a=-\alpha (s) = - 1
-\frac{s}{2}, \ b=-\alpha (t), \ c=-\alpha (u)$ with the
conditions $s+t+u = -8$ and $a+b+c=1$. According to \cite{freund1}
$p$-adic Veneziano amplitude is a simple $p$-adic counterpart of
(\ref{1.1}), i.e.
\begin{equation}
  A_p(a,b)        =  \int_{{\mathbb Q}_p} \vert x \vert_p^{a-1}
  \, \vert 1-x\vert_p^{b-1}\,  dx ,
     \label{1.2}
\end{equation}
where now $x \in  {\mathbb Q}_p$. In both (\ref{1.1}) and
(\ref{1.2}) kinematical variables $a, b, c$ are real (or
complex-valued) parameters. Thus in (\ref{1.2}) only string
world-sheet boundary $x$ is treated as $p$-adic variable, and all
other quantities maintain their usual  real values. Unfortunately,
there is a problem to extend the above product formula to the
higher-point functions. Some possibilities to construct $p$-adic
superstring amplitudes are considered in \cite{arefeva1} (see also
\cite{brekke1}, \cite{ruelle1},  and \cite{vladimirov2}). It seems
that to make further progress towards formulation of $p$-adic and
adelic superstring theory one has previously to develop
systematically the corresponding superalgebra and superanalysis.

 A promising  recent research in $p$-adic string theory has been mainly
related to an extension of adelic quantum mechanics
\cite{dragovich1}, \cite{dragovich2} (see also \cite{dragovich5})
and $p$-adic path integrals to string amplitudes \cite{dragovich4}
and quantum field theory \cite{dragovich3}. Also an effective
nonlinear $p$-adic string theory (see, e.g. \cite{brekke}) with an
infinite number of space and time derivatives has been recently of
a great interest in the context of the tachyon condensation
\cite{sen} (for a recent review, see \cite{freund}).

It is also worth mentioning successful formulation and development
of $p$-adic and adelic quantum cosmology (see \cite{dragovich6}
and references therein) which demonstrate discreteness of
minisuperspace with the Planck length $\ell_0$ as the elementary
one. There are also many other models in classical and quantum
physics, as well as in some related fields of other sciences,
which use $p$-adic numbers and adeles (for a recent activity, see
e.g. proceedings of conferences in $p$-adic mathematical physics
\cite{conference1}, \cite{conference2}).

The main mathematical motivations for employment of $p$-adic
numbers and/or adeles in modern mathematical physics are based on
the following facts: (i) the field of rational numbers ${\mathbb
Q}$ contains all experimental and observational numerical data;
(ii) ${\mathbb Q}$ is a dense subfield not only in the field of
real numbers ${\mathbb R}$ but also in all fields of $p$-adic
numbers ${\mathbb Q}_p$; (iii) ${\mathbb R}$  and ${\mathbb Q}_p$,
for all prime numbers $p$, exhaust all possible completions of
${\mathbb Q}$; and (iv) the local-global (Hasse-Minkowski)
principle, which states that, usually, if something is valid on
all local fields (${\mathbb R}$ and  ${\mathbb Q}_p$) then the
same is also valid on the global field (${\mathbb Q}$). One of the
main physical motivations is related to the well-known uncertainty
\begin{equation}
\Delta x \geq \ell_0 = \sqrt {\frac{\hbar G}{c^3}} \approx
10^{-33} cm, \label{1.3}
\end{equation}
which comes from some quantum gravity considerations.Accordingly
one cannot measure distances smaller than the Planck length
$\ell_0$. Since the derivation of (\ref{1.3}) is based on the
general assumption that real numbers and archimedean geometry are
valid at all scales it means that the usual approach is broken and
cannot be extended beyond the Planck scale without adequate
modification which contains non-archimedean geometry. The very
natural modification is to use adelic approach, since it contains
real and $p$-adic numbers which make all possible completions of
the number field ${\mathbb Q}$. As a next step it is natural to
consider possible relations between adelic and supersymmetry
structures.

In the sequel of this article we briefly review basic properties
of $p$-adic numbers, adeles and their functions, and then consider
$p$-adic and adelic superanalysis, which are important for
construction and investigation of the remarkable supersymmetric
models.

\section{$p$-Adic numbers, their algebraic extensions and adeles}

We review here some introductory notions on $p$-adic numbers,
their quadratic and algebraic extensions, and adeles. For further
reading one can use \cite{schikhof}, \cite{vladimirov1},
\cite{gelfand} and \cite{brekke}.

It is worth to start recalling  that the first infinite set of
numbers we encounter is the set ${\mathbb N}$ of natural numbers.
To have a solution of the simple linear equation $ x + a = b$ for
any $a, b \in {\mathbb N}$, one has to extend ${\mathbb N}$ and to
introduce the set ${\mathbb Z}$ of integers. Requiring that there
exists solution of the linear equation $ n x = m$ for any $0\neq
n, m \in {\mathbb Z}$ one obtains the set ${\mathbb Q}$ of
rational numbers. Evidently these sets satisfy ${\mathbb N}\subset
{\mathbb Z}\subset {\mathbb Q}$. Algebraically ${\mathbb N}$ is a
semigroup,  ${\mathbb Z}$ is a ring, and ${\mathbb Q}$ is a field.

To get ${\mathbb Q}$ from ${\mathbb N}$ only algebraic operations
are used, but to obtain the field ${\mathbb R}$ of real numbers
from ${\mathbb Q}$ one has to employ the absolute value which is
an example of the norm (valuation) on ${\mathbb Q}$. Let us recall
that a norm on ${\mathbb Q}$ is a map $||\cdot ||: {\mathbb Q}\to
{\mathbb R_{+}} = \{ x\in {\mathbb R}\,|\,\,  x\geq 0 \} $ with
the following properties: (i) $||x|| =0 \leftrightarrow x=0$, (ii)
$||x\cdot y|| = ||x||\cdot ||y|| $, and $\, ||x +y ||\, \leq \,
||x|| + ||y|| \,$ for all $x, y \in {\mathbb Q}$. In addition to
the absolute value, for which  we  use usual arithmetic notation
$|\cdot|_\infty$, one can introduce on ${\mathbb Q}$ a norm with
respect to each prime number $p$. Note that, due to the
factorization of integers, any rational number can be uniquely
written as $x = p^\nu \, \frac{m}{n}$, where $p,\, m,\, n$ are
mutually prime and $\nu \in {\mathbb Z}$. Then by definition
$p$-adic norm (or, in other words, $p$-adic absolute value) is $|
x|_p = p^{-\nu}$ if $ x \neq 0$ and $|0|_p =0$. One can verify
that $|\cdot |_p$ satisfies all the above conditions and moreover
one has strong triangle inequality, i.e. $|x + y|_p \leq \,  max
\, (|x|_p,\, |y|_p) $. Thus $p$-adic norms belong to the class of
non-archimedean (ultrametric) norms. There is only one
inequivalent $p$-adic norm for every prime $p$. According to the
Ostrowski theorem any nontrivial norm on ${\mathbb Q}$ is
equivalent either to the $|\cdot|_\infty$ or to one of the
$|\cdot|_p$. One can easily show that $|m|_p \leq 1$ for any $m
\in {\mathbb Z}$ and any prime $p$. The $p$-adic norm is a measure
of divisibility of the integer $m$ by prime $p$: the more
divisible, the $p$-adic smaller.  By Cauchy sequences of rational
numbers one can make completions of ${\mathbb Q}$ to obtain
${\mathbb R} \equiv {\mathbb Q}_\infty$ and the fields ${\mathbb
Q}_p$ of $p$-adic numbers using norms $|\cdot|_\infty$ and
$|\cdot|_p \,$, respectively. The cardinality of ${\mathbb Q}_p$
is the continuum as that one of ${\mathbb Q}_\infty$. $p$-Adic
completion of ${\mathbb Z}$ gives the ring ${\mathbb Z}_p = \{ x
\in {\mathbb Q}_p |\,\, |x|_p \leq 1 \}$ of $p$-adic integers.
Denote by ${\mathbb U}_p = \{ x \in {\mathbb Q}_p |\,\, |x|_p = 1
\}$ multiplicative group of $p$-adic units.

Any $p$-adic number $0 \neq x \in {\mathbb Q}_p$ has unique
representation (unlike real numbers) as the sum of a convergent
series of the form
\begin{equation}
x = p^\nu \, (x_0 + x_1 p + x_2 p^2 + \cdots + x_n p^n + \cdots )
, \quad \nu \in {\mathbb Z} , \quad x_n \in \{0, 1, \cdots, p-1 \}
. \label{2.1}
\end{equation}
It resembles representation of a real number $y= \pm\, 10^\mu \,
\sum_{k=0}^{-\infty} b_k 10^k ,\, \, \mu \in {\mathbb Z} , \, \,
b_k \in \{0, 1, \cdots, 9  \}$ , but in a sense with expansion in
the opposite way. If $\nu \geq 0$, then $x \in {\mathbb Z}_p$.
When $\nu = 0$ and $x_0 \neq 0$ one has $x \in {\mathbb U}_p$. Any
negative integer can be easily presented starting from the
representation for $- 1$:
\begin{equation}
- 1 = p-1 \ +\, (p-1) p\,  + \, (p-1) p^2 + \cdots + (p-1) p^n +
\cdots  . \label{2.2}
\end{equation}
Validity of (\ref{2.2}) can be shown by elementary arithmetics,
which is the same as in the real case, or treating it as the
$p$-adic convergent geometric series.

By the analogy with the real case, one uses the norm $|\cdot|_p$
to introduce $p$-adic metric $d_p (x, y) = |x - y|_p $, which
satisfies all necessary properties of metric with strengthened
triangle inequality in the non-archimedean (ultrametric) form:
$d_p (x, y) \leq max \,(\, d_p (x,z), \, d_p (z, y)\,)$. We can
regard $d_p (x,y)$ as a distance between $p$-adic numbers $x$ and
$y$. Using this (ultra)metric, ${\mathbb Q}_p$ becomes ultrametric
space and one can investigate the corresponding topology.
 Because of ultrametricity, the $p$-adic spaces have some exotic (from the
real point of view) properties and usual illustrative examples
are: a) any point of the ball $B_\mu (a) = \{x \in {\mathbb Q}_p\,
| \, \,\, |x - a|_p \leq p^\mu \}$ can be taken as its center
instead of $a$; b) any ball can be regarded as a closed as well as
an open set; c) two balls may not have partial intersection, i.e.
they are disjoint sets or one of them is a subset of the other;
and c) all triangles are isosceles. ${\mathbb Q}_p$ is
zerodimensional and totally disconnected topological space.
${\mathbb Z}_p$ is compact and ${\mathbb Q}_p$ is locally compact
space.

$ p\, {\mathbb Z}_p$ is a principal and the unique maximal ideal
of ${\mathbb Z}_p$. The corresponding residue field is the
quotient ${\mathbb Z}_p / p\, {\mathbb Z}_p$ , which is the Galois
field ${\mathbb F}_p$ with $p$ elements.

Recall that the field ${\mathbb C}$ of complex numbers can be
constructed as quadratic extension of ${\mathbb R}$ by using
formal solution $x = \sqrt {-1}$ of the equation $x^2 + 1 =0$ and
denoted by ${\mathbb C} = {\mathbb R}\, (\sqrt{-1})$. All elements
of ${\mathbb C}$ have the form $z = x + \sqrt{-1}\, y$ with $x, y
\in {\mathbb R}$.  ${\mathbb C}$ is algebraically closed,
metrically complete field, and a two-dimensional vector space.

Algebraic extensions of ${\mathbb Q}_p$ also exist and have more
complex structure. Quadratic extensions have the form ${\mathbb
Q}_p (\sqrt{\tau})$ with elements $z = x + \sqrt{\tau}\, y$ ,
where $x, y, \tau \in {\mathbb Q}_p$ and $\tau $ is not square
element of ${\mathbb Q}_p$ . For $p \neq 2$ there are three
inequivalent quadratic extensions and one can take $\tau =
\varepsilon, \, \varepsilon p, \, p$ , where $ \varepsilon =
\sqrt[p-1]{1} \, \in {\mathbb Q}_p$. When $p = 2$ there are seven
inequivalent quadratic extensions which may be characterized by
$\tau = -1, \pm 2, \pm 3, \pm 6$. Quadratic extensions are
complete but not algebraically closed. For solution of any higher
order algebraic equation one has to introduce at least one
extension. Namely, the equation $x^n - p = 0$ has solution $x =
\sqrt[n]{p}$ with the norm $|x|_p = p^{-\frac{1}{n}}$, which
exponent is a rational and not an integer number. Algebraic closer
of ${\mathbb Q}_p$, denoted by $\bar{\mathbb Q}_p$, is an infinite
dimensional vector space over ${\mathbb Q}_p$ which is not
complete. Completion of $\bar{\mathbb Q}_p$ gives ${\mathbb C}_p$
which is algebraically closed and metrically complete.

Real and $p$-adic numbers are continual extrapolations of rational
numbers along all possible notrivial and inequivalent metrics. To
consider real and $p$-adic numbers simultaneously and on equal
footing one uses concept of adeles. An adele $x$ (see, e.g.
\cite{gelfand}) is an infinite sequence $
  x= (x_\infty, x_2, \cdots, x_p, \cdots), $
where $x_\infty \in {\mathbb R}$ and $x_p \in {\mathbb Q}_p$ with
the restriction that for all but a finite set $\mathcal P$ of
primes $p$ one has  $x_p \in {\mathbb Z}_p $. Componentwise
addition and multiplication endow the ring structure to the set of
all adeles ${\mathbb A}$ , which is the union of restricted direct
products in the following form:
\begin{equation}
 {\mathbb A} = \bigcup_{{\mathcal P}} {\mathbb A} ({\mathcal P}),
 \ \ \ \  {\mathbb A} ({\mathcal P}) = {\mathbb R}\times \prod_{p\in
 {\mathcal P}} {\mathbb Q}_p
 \times \prod_{p\not\in {\mathcal P}} {\mathbb Z}_p \, .         \label{2.3}
\end{equation}

A multiplicative group of ideles ${\mathbb I}$ is a subset of
${\mathbb A}$ with elements $x= (x_\infty, x_2, \break \cdots,
x_p, \cdots)$ ,  where $x_\infty \in {\mathbb R}^\ast = {\mathbb
R} \setminus \{ 0\}$ and $x_p \in {\mathbb Q}^\ast_p = {\mathbb
Q}_p \setminus \{0 \}$ with the restriction that for all but a
finite set $\mathcal P$  one has that  $x_p \in {\mathbb U}_p$ .
Thus the whole set of ideles is
\begin{equation}
 {\mathbb I} = \bigcup_{{\mathcal P}} {\mathbb I} ({\mathcal P}),
 \ \ \ \ {\mathbb I} ({\mathcal P}) = {\mathbb R}^{\ast}\times \prod_{p\in {\mathcal P}}
 {\mathbb Q}^\ast_p
 \times \prod_{p\not\in {\mathcal P}} {\mathbb U}_p \, .         \label{2.4}
\end{equation}

A principal adele (idele) is a sequence $ (x, x, \cdots, x,
\cdots) \in {\mathbb A}$ , where $x \in  {\mathbb Q}\quad (x \in
{\mathbb Q}^\ast = {\mathbb Q}\setminus \{ 0\})$. ${\mathbb Q}$
and ${\mathbb Q}^\ast$ are naturally embedded in  ${\mathbb A}$
and ${\mathbb I}$ , respectively.

Let us define an ordering on the set ${\mathbb P}$, which consists
of all finite sets ${{\mathcal P}_i}$ of primes $p$, by
${{\mathcal P}_1} \prec {\mathcal P}_2$ if ${{\mathcal P}_1}
\subset {\mathcal P}_2$. It is evident that ${\mathbb A}({\cal
P}_1)\subset {\mathbb A}({\cal P}_2)$ when ${\mathcal P}_1 \prec
{\mathcal P}_2$. Spaces ${\mathbb A}({\cal P})$ have natural
Tikhonov topology and adelic topology in ${\mathbb A}$ is
introduced by inductive limit: $ {\mathbb A} = \lim
\mbox{ind}_{{\mathcal P} \in {\mathbb P}} {\mathbb A}({\cal P})$.
A basis of adelic topology is a collection of open sets of the
form $ W ({\mathcal P}) = {\mathbb V}_\infty \times \prod_{p \in
{\mathcal P}} {\mathbb V}_p\, \times \prod_{p \not \in {\mathcal
P}} {\mathbb Z}_p \, $, where ${\mathbb V}_\infty$ and ${\mathbb
V}_p$ are open sets in ${\mathbb R}$ and ${\mathbb Q}_p$ ,
respectively. Note that adelic topology is finer than the
corresponding Tikhonov topology. A sequence of adeles $a^{(n)}\in
{\mathbb A}$ converges to an adele $a \in {\mathbb A}$ if ${\it
i})$ it converges to $a$ componentwise and ${\it ii})\, $  if
there exist a positive integer $N$ and a set ${\mathcal P}$ such
that $\, a^{(n)}, \, a \in {\mathbb A} ({\mathcal P})$ when $n\geq
N$. In the analogous way, these assertions hold also for idelic
spaces ${\mathbb I}({\cal P})$ and ${\mathbb I}$. ${\mathbb A}$
and ${\mathbb I}$ are locally compact topological spaces.


\section{$p$-Adic and adelic analysis}

${\mathbb R} ,  {\mathbb Q}_p ,  {\mathbb C} ,  {\mathbb Q}_p
(\sqrt{\tau}) ,  {\mathbb C}_p ,  {\mathbb A} ,  {\mathbb I} $,
and the higher $p$-adic algebraic extensions,  form a large
environment for realization of various mappings and the
corresponding analyses. However only some of them have been used
in modern mathematical physics. Thus, in addition to the classical
real and complex analysis, the most important ones  are related to
the following mappings: $ ({\it i})  \, {\mathbb Q}_p \,\to
\,{\mathbb Q}_p \, , \, \,  ({\it ii}) \, {\mathbb Q}_p \,\to
\,{\mathbb C} , \,\,
 ({\it iii}) \, { \mathbb A} \,\to \,{\mathbb A} , \,\,
 ({\it iv}) \, { \mathbb A} \,\to \,{\mathbb C} , \,\,
 ({\it v}) \, { \mathbb Q}_p \,\to \,{\mathbb Q}_p (\sqrt{\tau}) \,$  and
 $\, ({\it vi})\,  {\, \mathbb Q}_p (\sqrt{\tau} ) \,\to \,{\mathbb C}$ .
We will give now some information about the corresponding
analyses.

$ Case \,  (i) \, {\mathbb Q}_p \,\to \,{\mathbb Q}_p $. All
functions from the real analysis which are given by infinite power
series $\sum a_n\, x^n$, where $a_n \in {\mathbb Q}$, can be
regarded also as $p$-adic if we take $x \in {\mathbb Q}_p $.
Necessary and sufficient condition for the convergence is $| a_n
\, x^n |_p  \to 0$ when $n \to \infty$.  For  example, $p$-adic
exponential function is
\begin{equation}
 \exp{ x} = \sum_{n=0}^{+\infty} \, \frac{x^n}{n !} \, , \label{3.1}
\end{equation}
where the domain of convergence is $| x |_p < | 2 |_p$. We see
that convergence is here bounded inside ${p\,\mathbb Z}_p$. Note
that
\begin{equation}
  | n! |_p = p^{- \frac{n - n'}{p -1}}\, \, , \label{3.2}
\end{equation}
where $n'$ is the sum of digits in the expansion of $n$ with
respect to $p$ , i.e. $n = n_0 + n_1 p + \cdots + n_k p^k$. An
interesting class of functions which domain of convergence is $
{\mathbb Z}_p $ has the form $ F_k (x) = \sum_{n \geq 0}  n! P_k
(n) x^n$ , where $P_k (n) = n^k  + C_{k-1} n^{k-1} + \cdots + C_0$
is a polynomial in $n$ with $C_i \in {\mathbb Z}$ (for various
properties of these functions, see \cite{dragovich7} and
references therein ).

Some functions can be constructed by the method of interpolation,
which is based on the fact that ${\mathbb N}$ is dense in
${\mathbb Z}_p$. Using the technique of interpolation $p$-adic
valued exponential, gamma and zeta functions are obtained.

In this case derivatives, antiderivatives and some definite
integrals are well defined. However there is a problem of
existence of the $p$-adic valued Lebesgue measure.

This kind of analysis is used in $p$-adic models of classical
physics.

$ Case\, ({\it ii}) \, {\mathbb Q}_p \,\to \,{\mathbb C} $. We
deal here with complex-valued functions of a $p$-adic argument.
Let us mention three important functions: the multiplicative
character $ \pi_s (x) = |x |_p^s \, , \,\, s\in {\mathbb C}$ ; the
additive character $\chi_p (x) = e^{2 \pi i \{ x\}_p}$ , where $\{
x\}_p $ is the fractional part of $x$ , i. e. $ \{ x\}_p  = p^{-
n} (x_0 + x_1 p + \cdots + x_{n-1} p^{n-1} )$ ; and the
characteristic function on ${\mathbb Z}_p$ which is
\begin{equation}
\Omega_p (|x|_p) = \left\{  \begin{array}{ll}
                 1,   &   |x|_p \leq 1,  \\
                 0,   &   |x|_p > 1 .
                 \end{array}    \right.  \label{3.3}
\end{equation}
Many other important functions may be obtained using these three
ones in some suitable ways.

Since ${\mathbb Q}_p$ and ${\mathbb Q}_p^\ast$ are locally compact
spaces there are on them the additive $dx$ and multiplicative
$d^\ast x$ Haar measures, respectively. With suitable
normalization, these measures have the following properties: $\,
d(x + a) = dx \,,\,\, a \in {\mathbb Q}_p \, ; \quad d(bx)= |b|_p
\, dx \,,\,\, b\in {\mathbb Q}_p^\ast \, ; \quad d^\ast (bx) =
d^\ast x \, , \,\, b\in {\mathbb Q}_p^\ast \, ;\quad  d^\ast x =
(1 - p^{-1}) \, |x |_p^{-1} \, dx \,. $ Integration with the Haar
measure is well defined. To overcome the problem with derivatives
one exploits approach with $p$-adic pseudodifferential operators
\cite{vladimirov1}.

The Gel'fand-Graev $p$-adic gamma and beta functions are:
\begin{equation}
\Gamma_p (a) = \int_{{\mathbb Q}_p} \, \chi_p (x)\, \pi_a (x) \,
|x |_p^{-1}\, dx \, = \, \frac{1 - p^{a -1}}{1 - p^{-a}} \,,
\label{3.4}
\end{equation}
\begin{equation}
 B_p (a, b) = \int_{{\mathbb Q}_p} \, \pi_a (x) \, |x |_p^{-1}
\, \pi_b (1-x) \, |1- x |_p^{-1} \, dx \, , \label{3.5}
\end{equation}
where $a, b \in {\mathbb R}$ or ${\mathbb C}$. This beta function
was used in construction of scattering amplitude (\ref{1.2}) for
$p$-adic open string (tachyon).

This kind of analysis is used also in $p$-adic quantum mechanics,
quantum cosmology and quantum field theory.

$Case \, ({\it iii}) \, { \mathbb A} \,\to \,{\mathbb A}$. This
case is an adelic collection of real and $p$-adic mappings which
enables to consider simultaneously and on equal footing real and
all $p$-adic aspects of a classical Lagrangian (and Hamiltonian)
system. In such case parameters for a given system should be
treated as rational numbers. Equations of motion must have an
adelic solution, i. e. function and its argument must have the
form of adeles.

$Case \, ({\it iv}) \, { \mathbb A} \,\to \,{\mathbb C} $. In this
case functions are complex-valued while their arguments are
adeles. The related analysis is used in adelic  approach to
quantum mechanics \cite{dragovich1}, \cite{dragovich2}, quantum
cosmology \cite{dragovich6}, quantum field theory
\cite{dragovich3} and string theory \cite{brekke},
\cite{vladimirov2}, \cite{dragovich4}. Many important
complex-valued functions from real and p-adic analysis can be
easily extended to this adelic case. Adelic multiplicative and
additive characters are:
\begin{equation}
 \pi_s (x) = |x|^s = |x_\infty|_\infty^s \, \prod_p \, |x_p|_p^s \,, \quad
 x\in {\mathbb I}\,, \quad s \in {\mathbb C} \,,  \label{3.6}
\end{equation}
\begin{equation}
\chi (x) = \chi_\infty (x_\infty)\, \prod_p \, \chi_p (x_p) = e^{-
2 \pi i x_\infty} \, \prod_p \, e^{2 \pi i \{ x_p\}_p}\,, \quad
x\in {\mathbb A} \,.    \label{3.7}
\end{equation}
Since all except finite number of factors in (\ref{3.6}) and
(\ref{3.7}) are equal to unity, it is evident that these infinite
products are convergent. One can show that $\pi_s (x) = 1$ if $x$
is a principal idele, and $\chi (x) = 1$ if $x$ is a principal
adele, i. e.
\begin{equation}
 |x|_\infty^s \, \prod_p \, |x|_p^s \, = 1 \,, \quad
 x\in {\mathbb Q}^\ast \,, \quad s \in {\mathbb C} \,,  \label{3.8}
\end{equation}
\begin{equation}
\chi_\infty (x)\, \prod_p \, \chi_p (x) = e^{- 2 \pi i x} \,
\prod_p \, e^{2 \pi i \{ x\}_p} \, =  1\,, \quad x\in {\mathbb Q}
\,. \label{3.9}
\end{equation}
It is worth noting that expressions (\ref{3.8}) and (\ref{3.9})
for $s=1$ represent the simplest adelic product formulas, which
clearly connect real and $p$-adic properties of the same rational
number. In fact, the formula (\ref{3.8}), for $s =1$, connects
usual absolute value and $p$-adic norms at the multiplicative
group of rational numbers ${\mathbb Q}^\ast$.

Maps $\varphi_{\mathcal P} : {\mathbb A}\to {\mathbb C}$, which
have the form
\begin{equation}
 \varphi_{\mathcal P} (x) = \varphi_{\infty} (x_\infty) \, \prod_{p \in {\mathcal P}}
 \varphi_p (x_p) \, \prod_{p \notin P} \Omega_p (|x_p|_p) \,,
 \label{3.10}
\end{equation}
where $\varphi_\infty (x_\infty)$ are infinitely differentiable
functions and fall to zero faster than any power of
$|x_\infty|_\infty$ as $|x_\infty|_\infty \to \infty$, and
$\varphi_p (x_p)$ are locally constant functions with compact
support, are called elementary functions on ${\mathbb A}$. All
finite linear combinations of the elementary functions
(\ref{3.10}) make the set ${\mathcal S}({\mathbb A})$ of
Schwartz-Bruhat functions $ \varphi (x)$.

${\mathbb A}$ is a locally compact ring and therefore there is the
corresponding  Haar measure, which is product of  the real and all
$p$-adic additive Haar measures. The Fourier transform of the
Schwartz-Bruhat functions $\varphi (x)$ is
\begin{equation}
\tilde{\varphi} (\xi) = \int_{\mathbb A} \varphi (x)\, \chi (x
\xi) \, dx    \label{3.11}
\end{equation}
and it maps ${\mathcal S}({\mathbb A})$ onto ${\mathcal
S}({\mathbb A})$. The Mellin transform of $\varphi (x)  \in
{\mathcal S}({\mathbb A})$ is defined using the multiplicative
character $|x|^s$ in the following way:
\begin{eqnarray}
\nonumber \Phi (s) =\int_{\mathbb I} \, \varphi (x) \, |x|^s \,
d^\ast x = \int_{\mathbb R} \varphi_\infty (x_\infty)\,
|x_\infty|_\infty^{s-1}\, dx_\infty \,\\
\times  \prod_p \,\int_{{\mathbb Q}_p} \varphi_p (x_p) \,
|x_p|_p^{s-1} \, \frac{dx}{1-p^{-1}}\, , \quad Re\, s > 1 .
\label{3.12}
\end{eqnarray}
$\Phi (s)$ may be analytically continued on the whole complex
plane, except $s=0$ and $s=1$, where it has simple poles with
residues $- \varphi (0)$ and $\tilde{\varphi} (0)$, respectively.
Denoting  by $\tilde{\Phi}$ the Mellin transform of
$\tilde{\varphi}$ then there is place the Tate formula
\cite{gelfand}
\begin{equation}
   \Phi (s) = \tilde{\Phi} (1 - s) \, .  \label{3.13}
\end{equation}
If we take $\varphi (x) = \sqrt[4]{2} \,\, e^{-\pi x_\infty^2} \,
\prod_p \Omega_p (|x_p|_p) $, which is the simplest ground state
of the adelic harmonic oscillator \cite{dragovich2}, then from the
Tate formula (\ref{3.13}) one gets the well-known functional
relation for the Riemann zeta function, i. e.
\begin{equation}
\pi^{- \frac{s}{2}} \, \Gamma \big( \frac{s}{2}\big) \, \zeta (s)
\, = \, \pi^{\frac{s -1}{2}} \, \Gamma \big( \frac{1 - s}{2}\big)
\, \zeta (1 - s) \,.    \label{3.14}
\end{equation}
We would like to emphasize this connection between the harmonic
oscillator and the Riemann zeta function.

In the last two cases, i. e. $({\it v}) \, { \mathbb Q}_p \,\to
\,{\mathbb Q}_p (\sqrt{\tau}) \,$  and
 $\, ({\it vi})\,  {\, \mathbb Q}_p (\sqrt{\tau} ) \,\to \,{\mathbb
 C}$,  $p$-adic quadratic extensions are used for values of
 functions and as values of the argument, respectively. The
 analysis
 for $({\it v}) \, { \mathbb Q}_p \,\to \,{\mathbb Q}_p (\sqrt{\tau})
\,$ is developed and used for a new type of non-archimedean
quantum mechanics (see, monograph  \cite{khrennikov} and
references therein ). Complex-valued functions of arguments from
${\, \mathbb Q}_p (\sqrt{\tau} )$ are also considered
\cite{gelfand} and employed for construction of the
Virasoro-Shapiro amplitudes for scattering of the $p$-adic closed
strings \cite{freund1}. Let us also mention that some mappings $
{\mathbb Q}_p (\sqrt{\tau}) \,\to \,{\mathbb Q}_p (\sqrt{\tau})$
have been used for investigation of classical dynamical systems
(see, book \cite{khrennikov1} and references therein)

\section{$p$-Adic and adelic superalgebra and superspace}

 As a next step to superanalysis we are going to review  here
real and $p$-adic superalgebra and superspace along approach
introduced by Vladimirov and Volovich \cite{vladimirov3},
\cite{vladimirov4} and elaborated  by Khrennikov
\cite{khrennikov2} (see also \cite{khrennikov}). Then, in the way
initiated by Dragovich \cite{dragovich8}, we shall generalize this
approach to the adelic case.

Let $ L ({\mathbb Q}_v ) = L_{0} ({\mathbb Q}_v)\, \oplus \, L_{1}
({\mathbb Q}_v)$ be $Z_2$-graded vector space over ${\mathbb Q}_v
, \, \, (v = \infty, 2, 3, \cdots, p, \cdots)$, where elements  $a
\in L_{0} ({\mathbb Q}_v)$ and $b\in L_{1} ({\mathbb Q}_v)$ have
even $(p(a) = 0)$ and odd $(p(b)=1)$ parities. Thus $L_{0}
({\mathbb Q}_v)$ and $L_{1} ({\mathbb Q}_v)$ are vector subspaces
of different parity. Such $L ({\mathbb Q}_v)$ space becomes
$v$-adic (i. e. real and $p$-adic) superalgebra, denoted by $
\Lambda ({\mathbb Q}_v ) = \Lambda_{0} ({\mathbb Q}_v)\, \oplus \,
\Lambda_{1} ({\mathbb Q}_v)$, if it is endowed by an associative
algebra with unity and multiplication with parity defined by
$p(ab) \equiv p(a) + p(b) \, (mod\, 2)$. Product of two elements
of the same (different) parity has even (add) parity.

Supercommutator is defined in the usual way: $[ a , b \} = a\, b -
(-1)^{p(a) p(b)} b\, a$. Superalgebra $\Lambda ({\mathbb Q}_v)$ is
called (super)commutative if $[ a , b \} =0$ for any $a \,, b $
which are elements of $ \Lambda_{0} ({\mathbb Q}_v)$ and $
\Lambda_{1} ({\mathbb Q}_v)$.

To  obtain a Banach space from the commutative superalgebra (CSA)
one has to introduce the corresponding norm
\begin{equation}
 || f g ||_v  \leq ||f||_v \, ||g||_v \,, \quad f, g \in \Lambda ({\mathbb
 Q}_v) ,
 \label{4.1}
\end{equation}
 which is at the end related to the absolute value $|\cdot|_\infty$ for the real case and to $p$-adic
norms $|\cdot|_p$ for $p$-adic cases.

As illustrative examples of commutative superalgebras one can
consider finite  dimensional $v$-adic Grassmann algebras $G (
{\mathbb Q}_v : \eta_1, \eta_2, \cdots, \eta_k)$ which dimension
is $2^k$ and generators $\eta_1, \eta_2, \cdots, \eta_k$ satisfy
anticommutative relations $ \eta_i \eta_j + \eta_j \eta_i = 0$.
These $\eta_i \, \eta_j$ can be realized as: 1) product of
annihilation operators $a_i \, a_j$ for fermions, 2) exterior
product $dx^i \, \wedge \, dx^j $ , and 3) as product of some
matrices. One can write
\begin{equation} G ( {\mathbb Q}_v : \eta_1,  \cdots,
\eta_k) = G_0 ( {\mathbb Q}_v : \eta_1,  \cdots, \eta_k) + G_1 (
{\mathbb Q}_v : \eta_1,  \cdots, \eta_k), \label{4.2}
\end{equation}
where $G_0 ( {\mathbb Q}_v : \eta_1, \eta_2, \cdots, \eta_k)$  and
$G_1 ( {\mathbb Q}_v : \eta_1, \eta_2, \cdots, \eta_k)$ contain
sums of $2^{k-1}$ terms with even and add number of algebra
generators $\eta_i$, respectively. Note that the role of commuting
and anticommuting coordinates will play these sums with even and
add parity and not the coefficients in  expansion over products of
$\eta_i$. As CSA one can also use the infinite dimensional
Grassmann algebra.

Let $\Lambda ({\mathbb Q}_v)$ be a fixed commutative $v$-adic
superalgebra. $v$-Adic superspace of dimension $(n,m)$ over
$\Lambda ({\mathbb Q}_v)$ is
\begin{equation}  {\mathbb Q}_{\Lambda
({\mathbb Q}_v)}^{n,m} = \Lambda_{0}^n ({\mathbb Q}_v) \, \times
\Lambda_{1}^m ({\mathbb Q}_v) \, ,\label{4.3} \end{equation} where
\begin{equation}
\Lambda_{0}^n ({\mathbb Q}_v) = \underbrace{\Lambda_{0} ({\mathbb
Q}_v)\times \, \cdots \,\times \, \Lambda_{0} ({\mathbb Q}_v)}_n
\, , \quad  \Lambda_{1}^m ({\mathbb Q}_v) =
\underbrace{\Lambda_{1} ({\mathbb Q}_v)\times \, \cdots \,\times
\, \Lambda_{1} ({\mathbb Q}_v)}_m \,.
   \label{4.4} \end{equation}
This superspace is an extension of the standard $v$-adic space,
which has now $n$ commuting and $m$ anticommuting coordinates.

 The points of the superspace ${\mathbb Q}_{\Lambda
({\mathbb Q}_v)}^{n,m}$ are
\begin{eqnarray}
\nonumber X^{(v)} = (X^{(v)}_1, X^{(v)}_2,\cdots, X^{(v)}_n,
X^{(v)}_{n+1}, \cdots, X^{(v)}_{n+m})\\ = (x^{(v)}_1, x^{(v)}_2,
\cdots, x^{(v)}_n, \theta^{(v)}_1, \cdots, \theta^{(v)}_m) =
(x^{(v)}, \theta^{(v)}), \label{4.5}
\end{eqnarray}
 where coordinates $x^{(v)}_1, x^{(v)}_2,
\cdots, x^{(v)}_n$ are commuting, with $p(x^{(v)}_i) =0$, and
$\theta^{(v)}_1, \break \theta^{(v)}_2, \cdots, \theta^{(v)}_m$
are anticommuting (Grassmann) ones, with $p(\theta^{(v)}_j) = 1$.
Since the supercommutator $[ X^{(v)}_i , X^{(v)}_j \} = X^{(v)}_i
X^{(v)}_j - (-1)^{p(X^{(v)}_i) p(X^{(v)}_j)} X^{(v)}_j X^{(v)}_i =
0$, the coordinates $X^{(v)}_i , \, \, (i = 1, 2, \cdots, n+m )$
are called supercommuting.

A norm of $X^{(v)}$ can be defined as
\begin{equation}
||X^{(v)}||_v =    \left\{  \begin{array}{ll}
                 \sum_{i=1}^n ||x_i^{(\infty)} ||_\infty + \sum_{j=1}^m ||\theta_j^{(\infty)} ||_\infty,   & v =\infty ,  \\
                 \mbox{max}_{1\leq i \leq n, \, 1\leq j \leq m}\, (\, ||x_i^{(p)} ||_p\,, ||\theta_j^{(p)}||_p \,) ,  &   v =p
                 \,,
                 \end{array}    \right.
 \label{4.6}
\end{equation}
where $||X^{(p)}||_p \,, \, ||x_i^{(p)}||_p \,\, \mbox{and} \,\,
||\theta_j^{(p)}||_p $ are non-archimedean norms. In the sequel,
to decrease number of indices we often omit some of them when they
are understood from the context.

We can now turn to the adelic case of superalgebra and superspace.
Let us start with the corresponding $Z_2$-graded  vector space
over ${\mathbb A}$ as
\begin{equation}
L ({\mathbb A}) = \bigcup_{\mathcal P} L ({\mathcal P}) \, , \,
\quad L ({\mathcal P})  = L ({\mathbb R}) \times \prod_{p \in
{\mathcal P}} L ({\mathbb Q}_p) \times \prod_{p \not\in {\mathcal
P}} L ({\mathbb Z}_p ) \, ,\label{4.7}
\end{equation}
where $ L ({\mathbb Z}_p )  = L_0 ({\mathbb Z}_p )\oplus L_1
({\mathbb Z}_p )$ is a graded  space over the ring of $p$-adic
integers $ {\mathbb Z}_p  $ (and ${\mathcal P}$ is a finite set of
primes $p$). Graded vector space (\ref{4.7}) becomes adelic
superalgebra
\begin{equation}
\Lambda ({\mathbb A}) = \bigcup_{\mathcal P} \Lambda ({\mathcal
P}) \, , \, \quad \Lambda ({\mathcal P})  = \Lambda ({\mathbb R})
\times \prod_{p \in {\mathcal P}} \Lambda ({\mathbb Q}_p) \times
\prod_{p \not\in {\mathcal P}} \Lambda ({\mathbb Z}_p ) \,
,\label{4.8}
\end{equation}
by requiring that $ \Lambda ({\mathbb R}) \, , \Lambda ({\mathbb Q
}_p ) \, , \,\mbox{and} \, \Lambda ({\mathbb Z}_p )  = \Lambda_0
({\mathbb Z}_p )\oplus \Lambda_1 ({\mathbb Z}_p ) $ are
superalgebras. Adelic supercommutator may be regarded as a
collection of real and all $p$-adic supercommutators. Thus adelic
superalgebra (\ref{4.8}) is commutative.  An example of
commutative adelic superalgebra is the following adelic Grassmann
algebra:
\begin{equation} G ({\mathbb A}: \eta_1, \eta_2, \cdots, \eta_k) =
\bigcup_{\mathcal P} G ({\mathcal P} :  \eta_1, \eta_2, \cdots,
\eta_k)\, \label{4.9}
\end{equation}
\begin{eqnarray}
 \nonumber G ({\mathcal P} : \eta_1,  \cdots, \eta_k) = G ({\mathbb R} :
\eta_1, \cdots, \eta_k) \\ \times  \prod_{p \in{\mathcal P}} G
({\mathbb Q}_p  : \eta_1,  \cdots, \eta_k)  \times \prod_{p
\not\in{\mathcal P}} G( {\mathbb Z}_p :\eta_1,  \cdots, \eta_k) .
\label{4.10}
\end{eqnarray}

Banach commutative superalgebra for $\Lambda ({\mathcal P})$
defined in (\ref{4.8}) obtains by  taking all of  $\Lambda
({\mathbb Q}_\infty) ,\, \Lambda ({\mathbb Q}_p) , \, \Lambda
({\mathbb Z}_p)$ to be Banach spaces. The $\Lambda ({\mathcal P})$
will have the corresponding Tikhonov topology. Then the
corresponding Banach adelic space is inductive limit $\Lambda
({\mathbb A}) = \lim \mbox{ind}_{{\mathcal P} \in {\mathbb P}}
\,\, \Lambda ({\mathcal P}) $, and in this way it  gets an adelic
topology.

Adelic superspace of dimension $(n,m)$ has the form
\begin{equation}
{\mathbb A}^{n,m}_{\Lambda ({\mathbb A})} = \bigcup_{\mathcal P}
{\mathbb A}^{n,m}_{\Lambda ({\mathbb A})} \,\big({\mathcal
P}\big)\, , \quad {\mathbb A}^{n,m}_{\Lambda ({\mathbb A})} \,
\big( {\mathcal P} \big) = {\mathbb R}_{\Lambda ({\mathbb
R})}^{n,m} \times \prod_{p\in {\mathcal P}} {\mathbb Q}_{\Lambda
({\mathbb Q}_p)}^{n,m}  \times \prod_{p\not\in {\mathcal P}}
{\mathbb Z}_{\Lambda ({\mathbb Z}_p)}^{n,m}\, , \label{4.11}
\end{equation}
where ${\mathbb Z}_{\Lambda ({\mathbb Z}_p)}^{n,m} $ is
$(n,m)$-dimensional $p$-adic superspace over Banach commutative
superalgebra ${\Lambda ({\mathbb Z}_p)}$.

Points $X$ of adelic superspace  have the  coordinate form $X =
(X^{(\infty)}, X^{(2)}, \cdots ,\break X^{(p)}, \cdots, )$, where
for all but a finite set of primes ${\mathcal P}$ it has to be
$||X^{(p)}||_p = \mbox{max}_{1\leq i \leq n, \, 1\leq j \leq m}\,
(\, ||x_i^{(p)} ||_p\,, ||\theta_j^{(p)} ||_p \,) \, \, \leq 1 $ ,
i. e. $x_i^{(p)} \, \in {\Lambda_0} ({\mathbb Z}_p)\,\, \mbox{and}
\,\, \theta_j^{(p)} \, \in {\Lambda_1} ({\mathbb Z}_p)$ .

\section{Elements of $p$-adic and adelic superanalysis}

Superanalysis is related to a map from one superspace to the
other. Since we have formulated here many superspaces which are
distinctly valued, therefore one can introduce many kinds of
mappings between them and one can get plenty of superanalyses. For
instance, one can consider the following cases: real
$\longrightarrow $ real , $p$-adic  $\longrightarrow $ $p$-adic ,
adelic $\longrightarrow $ adelic, real $\longrightarrow $ complex
, $p$-adic $\longrightarrow $ complex , and adelic
$\longrightarrow $ complex .  In the sequel we will restrict our
consideration  to the cases without complex-valued functions. In
fact we will investigate two types of maps:
\begin{equation}
 F_v :\, V_v \to V'_v , \quad \quad \Phi_{\mathbb A} :\, W ({\mathcal P})
\to W' ({\mathcal P}') \,, \label{5.1}
\end{equation}
where $V_v\,   \subset  {\mathbb Q}_{\Lambda ({\mathbb
Q}_v)}^{n,m} $ ,  $\,\, V'_v\,   \subset  {\mathbb Q}_{\Lambda
({\mathbb Q}_v)}^{n,m} $ , and $\, W ({\mathcal P})   \subset
{\mathbb A}_{\Lambda ({\mathbb A})}^{n,m} ( {\mathcal P}) $ ,
$\,\, W' ({\mathcal P}')   \subset {\mathbb A}_{\Lambda ({\mathbb
A})}^{n,m} ( {\mathcal P}') $.

{\it Case} $F_v$. The function $F_v (X)$ is continuous  in the
point $X \in V_v $  if
\begin{equation}
\lim_{||h||_v \to 0}  || F_v (X + h) - F_v ||_v  = 0 \,,
\label{5.2}
\end{equation}
and it is continuous in $V_v$ if (\ref{5.2}) is satisfied for all
$X \in V_v$. This function $F_v$ is superdifferentiable in $X \in
V_v $ if it can be presented as
\begin{equation}
F_v (X + h) = F_v (X) + \sum_{i=1}^{n+m}\, f_i (X) \, h_i + g(X,
h)\,, \label{5.3}
\end{equation}
where $f_i (X) \in V'_v$ and $|| g (X,h)||_v \, ||h||_v^{-1} \to
0$ when $||h||_v \to 0$. Then  $f_i (X)$ are called partial
derivatives of $F_v$ in the point $X$ with respect to $X_i$ and
denoted by
\begin{equation}
f_i (X) = \frac{\partial F_v (X)}{\partial X_i} = \frac{\partial
F_v (x, \theta)}{\partial x_i} \,, \,\,\,  f_{n +j} (X) =
\frac{\partial F_v (X)}{\partial X_{n+j}} = \frac{\partial F_v (x,
\theta)}{\partial \theta_j} \,,  \label{5.4}
\end{equation}
where $i = 1, 2, \cdots, n$  and $j = 1, 2, \cdots, m$ . The
superdifferential is
\begin{equation}
D F_v (X)  = \sum_{i=1}^{n+m}\, \frac{\partial F_v (X)}{\partial
X_i} \, h_i . \label{5.5}
\end{equation}
If  $F_v$ is an $(n+m)$-component function then partial
derivatives form $(n+m)\times (n+m)$ Jacobi matrix. The above
introduced derivatives are known as the right ones. One can also
introduce the left derivatives by change $f_i (X) \, h_i \to h_i
\, f_i (X) $  in (\ref{5.3}). Higher order derivatives can be
introduced in the analogous way. Note that partial derivatives
with odd coordinates anticommute: $\frac{\partial}{\partial
\theta}_i \, \frac{\partial}{\partial \theta}_j  = - \,
\frac{\partial}{\partial \theta}_j \, \frac{\partial}{\partial
\theta}_i $ . Since in this approach coordinates $x_i$ and
$\theta_j$ are composed of coefficients in commutative
superalgebra ${\Lambda ({\mathbb Q}_v)}$ there exist the
corresponding Cauchy-Riemann conditions (for details, see,
\cite{vladimirov3}).

Note that superdifferentiability is closely related to the
Frech$\grave{e}$t differentiability in the Banach spaces.

The corresponding integral calculus is based on appropriately
constructed  differential forms \cite{vladimirov4}. Integration
over noncommuting variables employs the standard rules
\begin{equation}
  \int d\theta = 0\,, \quad  \int \theta\,  d\theta = 1 \,.
   \label{5.6}
\end{equation}

When one of the commuting coordinates $x_i$ is the time and the
others are spatial in the superspace $ {\mathbb Q}_{\Lambda
({\mathbb Q}_v)}^{n,m} $, the functions are called superfields in
supersymmetric physical models.

Various aspects of $p$-adic superanalysis have been considered in
detail and many of them can be found in the papers
\cite{khrennikov3}, \cite{khrennikov4}, \cite{khrennikov5},
\cite{khrennikov6}, \cite{khrennikov7}.

The corresponding adelic valued functions (superfields)
$\Phi_{\mathbb A}$ must satisfy adelic structure, i.e.
$\Phi_{\mathbb A} (X) = (F_\infty\, (X^{(\infty)}) ,\,  F_2 \,
(X^{(2)}) ,\,  \cdots ,\, F_p \, (X^{(p)}) , \cdots )$ with
condition $||F_p||_p \, \leq 1$ for all but a finite set of primes
${\mathcal P}$. According to this adelic property and the above
$v$-adic superanalysis one obtains the corresponding adelic
superanalysis.

\section{Simple model of superanalysis with Grassmann algebra}

According to the above approach, essential properties of
superanalysis should not depend on the concrete choice of a
$Z_2$-graded commutative Banach algebra. Hence, to realize a
simple and instructive model of superanalysis it is natural to
make construction over the Grassmann algebra with two
anticommuting elements (generators).

Let $\eta_1 , \eta_2$ be two fixed generators of the $v$-adic
Grassmann algebra $G ({\mathbb Q}_v : \eta_1 , \eta _2 )$ , i. e.
$\eta_i $ satisfy $\eta_1 \eta_2 = - \eta_2  \eta_1$ and $\eta_1
\eta_1  = \eta_2 \eta_2  = 0$. Then $G (\mathbb{Q}_v : \eta_1 ,
\eta_2 )$ can be presented as (super)commutative superalgebra
\begin{equation}
G (\mathbb{Q}_v : \eta_1 , \eta_2 ) = \Lambda (\mathbb{Q}_v) =
\Lambda_0 (\mathbb{Q}_v) \oplus \Lambda_1 (\mathbb{Q}_v) .
\label{6.1}
\end{equation}
 The
corresponding elements $x \in \Lambda_0 (\mathbb{Q}_v)$ and $
\theta \in \Lambda_1 (\mathbb{Q}_v)$ have the form
\begin{equation}
x = u + v \, \eta_1 \, \eta_2 \,, \quad  \theta = \alpha \, \eta_1
+ \beta \, \eta_2 \,, \label{6.2}
\end{equation}
 where $u, \, v ,\, \alpha,\, \beta
\in \mathbb{Q}_v$. According to the above approach to
superanalysis (see \cite{vladimirov3} and \cite{vladimirov4} )
these $x$ and $\theta$ become commuting and anticommuting
$v$-adic variables (coordinates), respectively. One can easily
verify that any two values $\theta_1 = \alpha_1 \, \eta_1  +
\beta_1 \, \eta_2 $ and $\theta_2 = \alpha_2 \, \eta_1 + \beta_2
\, \eta_2 $  are anticommuting, i.e. $\theta_i \, \theta_j +
\theta_j \, \theta_i =0, \quad i,j = 1,2$. Since the parity  $p
(\eta_1 ) = p (\eta_2) = 1$ and $p (u_i) = p (v_i) = p (\alpha_i)
= p (\beta_i) = 1$ it follows that $p (x_1) = p (x_2) = 0$ and $p
(\theta_1) = p (\theta_2) = 1$.

The corresponding real and p-adic norms of $x $ and $\theta$ in
(\ref{6.2}) are as follows:
\begin{equation} || x||_v =  \left\{
\begin{array}{ll} |u|_\infty  +  |v|_\infty \,, &  v= \infty \\
\mbox{max} \{|u|_p \,, |v|_p  \} \, , & v =p \,,
  \label{6.3}
\end{array} \right.
\end{equation}
\begin{equation} || \theta||_v =  \left\{
\begin{array}{ll} |\alpha|_\infty  +  |\beta|_\infty \,, &  v= \infty \\
\mbox{max} \{|\alpha|_p \,, |\beta|_p  \} \, , & v =p \,,
  \label{6.4}
\end{array} \right.
\end{equation}
where $|\cdot|_\infty$ and $|\cdot|_p$ are  usual norms of real
and $p$-adic numbers. With these norms (\ref{6.3}) and
(\ref{6.4}),
 $ \Lambda (\mathbb{Q}_\infty)$  and $\Lambda (\mathbb{Q}_p)$ become
 Banach superalgebras.

Now one can introduce  superspace  (\ref{4.3}) with $n$  commuting
and $m$ anticommuting coordinates
\begin{equation}
x_i = u_i + v_i \, \eta_1 \, \eta_2\,, \quad \theta_j = \alpha_j
\, \eta_1 + \beta_j \, \eta_2 \,, \quad i= 1,  \cdots , n \,, \,\,
j = 1,  \cdots , m\,. \label{6.5}
\end{equation}

According to (\ref{4.6}), (\ref{6.3}) and (\ref{6.4}) the norm of
$v$-adic superspace points $X^{(v)} = ( x^{(v)}\,, \theta^{(v)})$
is
\begin{equation} || X^{(v)}||_v =  \left\{
\begin{array}{ll} \sum_{i=1}^n \,( |u_i|_\infty  +  |v_i|_\infty ) +
\sum_{j=1}^m (|\alpha_j|_\infty  + |\beta_j|_\infty ) \,, &  v= \infty \\
\mbox{max}_{i, j} \,\, \{\, |u_i|_p \,, |v_i|_p \,, |\alpha_j |_p
\,, |\beta_j |_p \, \} \, , & v =p \,,
  \label{6.6}
\end{array} \right.
\end{equation}

In construction of adelic superspace (\ref{4.11}) one has to take
care that $\mbox{max}_{i, j} \,\, \{ \, |u_i|_p \,, \break |v_i|_p
\,, |\alpha_j |_p \,, |\beta_j|_p \, \} \leq  1 $ for all
possibilities of finite set ${\mathcal P}$.

\section{Conclusion}

In this article we have given a brief review of real and $p$-adic
analysis and superanalysis on the Banach commutative superalgebra.
This is approach which generalizes analysis of complex functions
with many complex variables. An introduction to adelic
superanalysis is also presented. Some elements of this approach
are illustrated using the Grassmann algebra of two anticommuting
generators. As a next step we plan to consider $p$-adic and adelic
superanalysis with complex-valued superfields, as well as to
develop adelic theory of supersymmetry and to construct $p$-adic
analogs and adelic models of superstring and M-theory.

%


\section*{Acknowledgments}

One of the authors (B.D.) would like to thank Prof. V. Dobrev for
invitation to participate in QTS-4 and give a plenary talk on 'VI
Workshop on Lie Theory and Its Applications in Physics', Varna
2005, as well as for the hospitality and stimulative scientific
atmosphere. Work on this article was partially supported by the
Serbian Ministry of Science and Environmental Protection under the
project no. 1426, and Grants: "EU-network: quantum probability and
applications" and "Swedish Royal Academy of Science for
collaboration with states of former Soviet Union".


\end{document}